\documentclass[graybox]{svmult}

\usepackage{type1cm}        
\usepackage{makeidx}         
\usepackage{graphicx}        
\usepackage{multicol}        
\usepackage{multirow}
\usepackage[percent]{overpic} 
\usepackage[bottom]{footmisc}
\usepackage{tikz} 
\usetikzlibrary{shapes.geometric, arrows}
\usepackage{pgfplots} 
\pgfplotsset{width=\linewidth, compat=1.6} 
\usepackage{pgfplotstable} 
\usepackage{subcaption}
\usepackage{float}
\usepackage{newtxtext}       %
\usepackage{newtxmath}       
\usepackage{array}
\usepackage{xcolor}
\usepackage[backend=biber, sorting=nyt, giveninits=true, style=numeric, doi=false, url=false, isbn=false, maxbibnames=99]{biblatex} 
\DeclareNameAlias{default}{family-given}
\DeclareNameAlias{sortname}{family-given}
\DeclareFieldFormat{labelnumberwidth}{#1.\adddot\space}

\DeclareFieldFormat[article]{journaltitle}{#1}
\renewbibmacro*{journal+issuetitle}{
  \usebibmacro{journal}
  \setunit*{\addspace}
  \usebibmacro{volume+number+eid}
  \setunit{\addcomma\space}
  \usebibmacro{issue+date}
  \newunit
}
\DeclareFieldFormat[inproceedings]{booktitle}{#1}
\renewbibmacro*{journal+issuetitle}{
  \usebibmacro{journal}
  \setunit*{\addspace}
  \usebibmacro{volume+number+eid}
  \setunit{\addcomma\space}
  \usebibmacro{issue+date}
  \newunit
}
\DeclareFieldFormat[article]{volume}{\textbf{#1}}\tikzset{every picture/.style={line width=0.5pt}}
\usepackage{siunitx}
\usepackage{hyperref}

\usepackage{etoolbox}
\AtBeginBibliography{\small\setlength{\itemsep}{0.5pt plus 0.6ex}}
\usepackage{booktabs}

\usepackage[normalem]{ulem} 

\makeindex             
\graphicspath{figures}

\usepackage{float}

\usepackage{subcaption}

\newcommand{\code}[1]{\texttt{#1}}

\definecolor{color1}{RGB}{230, 25, 75}
\definecolor{color2}{RGB}{60, 180, 75}
\definecolor{color3}{RGB}{255, 225, 25}
\definecolor{color4}{RGB}{0, 130, 200}
\definecolor{color5}{RGB}{245, 130, 48}
\definecolor{color6}{RGB}{70, 240, 240}
\definecolor{color7}{RGB}{240, 50, 230}
\definecolor{color8}{RGB}{250, 190, 212}
\definecolor{color9}{RGB}{0, 128, 128}
\definecolor{color10}{RGB}{220, 190, 255}
\definecolor{color11}{RGB}{170, 110, 40}
\definecolor{color12}{RGB}{255, 250, 200}
\definecolor{color13}{RGB}{128, 0, 0}
\definecolor{color14}{RGB}{170, 255, 195}
\definecolor{color15}{RGB}{0, 0, 128}
\definecolor{color16}{RGB}{128, 128, 128}
\definecolor{color17}{RGB}{255, 255, 255}

\definecolor{gray1}{gray}{0.1}
\definecolor{gray2}{gray}{0.3}
\definecolor{gray3}{gray}{0.5}
\definecolor{gray4}{gray}{0.7}
\definecolor{gray5}{gray}{0.9}
\begin{document}

\title*{The Role of Interfacial Tension in Direct Numerical Simulations of Drop-Film Interaction for Immiscible Fluids}
\author{Rishi Dhar, David Gösele, Paul Saumet, Bernhard Weigand and Kathrin Schulte}
\authorrunning{R. Dhar, D. Gösele, P. Saumet, B. Weigand and K. Schulte}
\institute{Rishi Dhar, David Gösele, Bernhard Weigand, Kathrin Schulte
\at Institute of Aerospace Thermodynamics (ITLR), University of Stuttgart, Pfaffenwaldring 31, 70569~Stuttgart, Germany, \email{rishi.dhar@itlr.uni-stuttgart.de} and \email{david.gösele@itlr.uni-stuttgart.de}
\and
Paul Saumet \at High Performance Computing Center, Stuttgart \email{hpcpsaum@hlrs.de}} 
%
%
\maketitle
\vspace{-12mm}
\abstract{
Immiscible fluid interactions are visible across industrial applications, where a drop-film interaction can be studied on a fundamental level. 
Many experimental studies have reported variations in the values of interfacial tension, which is a critical component for numerical investigations. Isolating all the geometric and fluid material parameters and varying the interfacial tension can be useful to check its influence. Numerical investigations using Free Surface 3D (FS3D), have been conducted using validated results to compare varying values of interfacial tension and evaluating the sensitivity. 
A grid independence study compared the compound crown height of a splash to determine the required resolution for validation. A qualitative validation showed FS3D could correctly capture the impact morphology while varying the viscosity ratio of drop and film liquid when compared to the experimental results. A quantitative validation for a water drop impacting onto an oil film shows a good match for the crown heights of the numerical and the experimental data. The same setup was then extended to study the variation of interfacial tension where the deviation of the overall compound crown height and spreading diameter of the internal crowns were compared. Results revealed minor changes in the compound crown height and spreading diameter of the drop liquid, but the internal crown composition showed significant differences. 
In order to run FS3D efficiently on the new supercomputer Hunter, which has a new APU architecture based system, extensive work had to be done.
To adapt to the new hardware architecture, large parts of FS3D have been ported to efficiently utilize the 
AMD Instinct™ MI300A accelerated processing units (APUs) at HLRS using OpenMP. 
Implementation of Umpire memory pools significantly improved performance for larger workloads per APU. 
The GPU-accelerated code achieves a 4 time speedup compared to CPU-only execution on the same hardware.
Strong and weak scaling tests have been conducted,
showing good strong scaling for up to 4 APUs,
and linear weak scaling for up to 512 APUs,
resulting in a total of $4096^3$ cells for the first time.

}

\newpage
\section{Introduction}
\label{sec:Introduction}
\vspace{-4mm}
\cite{Mathi123}Immiscible fluid interactions are prevalent across various industrial applications, including chemical reactors \cite{Lakatos2015}, oil-spill remediation \cite{Li2016}, and cosmetic formulations \cite{Yukuyama2015}. These processes often involve complex spray–film interactions, which are challenging to analyze due to their transient and multiphase nature. On a fundamental level, a drop-film interaction can be studied for evaluating the macro parameters of a spray such as splashing thresholds and spreading dimensions. For miscible fluid systems, drop–film interactions typically evolve through four primary stages \cite{tuprints8986}: (1) the initial impact and contact between the droplet and the liquid film, (2) the spreading of the droplet accompanied by the formation of a crown, and (3) rim destabilization at the crown’s apex (4) formation of secondary droplets and crown collapse. Investigations suggest that during the initial phases of impact of drop on film and crown formation, outcomes for miscible fluid interactions are significantly influenced by the film height ($h_{\textnormal{film}}$) and droplet velocity ($U_{\textnormal{drop}}$) \cite{Chen2017}. For drop-film interactions of immiscible fluids \cite{Che2018,Shaikh2017,Kittel2018} the viscosity ratio of the drop and film liquid and the surface tension along with the interfacial tension become critical in determining the output characteristics \cite{qin2023}. \\
While the individual properties of each liquid such as viscosity and surface tension to describe miscible fluid interactions, immiscible systems require consideration of interfacial tension as a parameter. Interfacial tension arises along the contact line where the two liquid phases meet and can significantly influence the dynamics of the interaction. Experimental studies on drop–film interactions involving immiscible fluids have demonstrated that interfacial tension plays a decisive role in governing the resulting flow behavior and interfacial stability \cite{Che2018}. Experimental results indicate a noticeable difference in the velocities of the crown tips formed by the drop and film liquids during immiscible interactions, which is attributed to the presence of interfacial tension \cite{Wu2021}. Interfacial tension is commonly measured using the pendant-drop method \cite{Berry2015}, a technique that relies heavily on the image resolution of the droplet profile. As a result, the accuracy of the measured values can vary significantly depending on the quality and precision of the imaging system, leading to variations. As an example, pure water has a surface tension of 72 mN/m and different silicon oils have a surface tension of 20 mN/m, but literature suggests varying values of interfacial tension at room temperature. Kittel \cite{tuprints8986} in her dissertation observed the interfacial tension to be around 51 mN/m for S10 silicon oil and water whereas \textcite{Shaikh2017} in his set of studies found the interfacial tension to be around 23-25 mN/m for water and silicon oils S5, S10, S50 and S100.  While such variations in the interfacial tension can be accounted to the impurities or presence of additional liquid in small percentages \cite{udeagbara2010,binks2000}, interfacial tension remains a critical input parameter for numerical investigations, raising questions regarding its sensitivity. \\
Isolating the effects of interfacial tension can be difficult in a drop-film interaction scenario considering the range of complexities observed in experimental studies during the first two stages. Large-scale experimental studies of droplet impacts reveal only geometrical features of splash morphologies, which account for the interplay of various fluid properties and impact conditions. If the impact conditions and fluid properties are kept constant, the viscosity ratio between the droplet and the film liquids can significantly influence the crown formation, highlighting the criticality of the selection of the drop and film liquid. Kittel et al. \cite{Kittel2018} observed that an oil drop impact over a water film adheres to the film liquid whereas exchanging the drop and film liquids lead to separate crowns formed by the drop and film liquid. Kittel et al. \cite{Kittel2018} hence incorporated the viscosity ratio for evaluating splashing thresholds, underscoring its importance in such interactions.\\
Studies isolating the viscosity ratio and focusing on the variations in interfacial tension are scarce. Investigating these aspects would not only enhance the understanding of interfacial dynamics but also provide valuable sensitivity analyses for numerical models. Such research could establish uncertainty thresholds, thereby improving the reliability of simulations and experimental interpretations in droplet impact studies.\\
The current study is a preliminary investigation on the effects of interfacial tension values and its role specifically in the  drop-film interaction. Section \ref{sec:description} provides an overview of the multiphase simulation code FS3D being used for carrying out the investigations. 
Section \ref{sec:simulation_results} highlights the setup for the numerical investigation of a drop-film interaction with immiscible drop and film liquids. A grid independence study reveals the resolution necessary to conduct a quantitative validation study. The effects of variation in viscosity ratio of drop and film liquid have been validated qualitatively. The grid suggested during the grid independence study has been used for a quantitative validation where the numerical values of crown height have been compared with the experimental results of \textcite{Che2018}. After reaching good agreement, the interfacial tension of the same set of liquids have been altered to check the effects observed within the impact morphology. \\
This study extensively utilized the computational nodes of the Hawk system. With the transition to Hunter—a newer system featuring an APU (Accelerated Processing Unit) architecture—significant modifications to the codebase were required to effectively leverage GPU computational resources. Section \ref{sec:performance} details the methodologies employed for GPU offloading and scaling, including the integration of Umpire, a memory management tool used to accelerate variable allocation. To evaluate performance, an oscillating droplet test case was employed.  
\vspace{-5mm}
\section{Mathematical Description and Numerical Approach}
\label{sec:description}
\vspace{-4mm}
Free Surface 3D (FS3D), a program package developed at ITLR, performs direct numerical simulation (DNS) for dynamic multiphase simulations \cite{Eisenschmidt2016}. It has been used for numerical investigations of different physical phenomenon such as drop-film interaction \cite{steigerwald2018}, jet atomization \cite{Ertl2017}, non-newtonian fluids \cite{Steigerwald2024} and immiscible drop-collision \cite{Potyka2023}. In order to resolve fine spatial and temporal scales, FS3D is highly parallelized using Message Passing Interface (MPI) and Open Multi-processing (OpenMP). 
FS3D uses finite volume method to solve discretized Cartesian meshes, whilst solving the mass and momentum equation of the incompressible Navier-Stokes equations
\begin{equation}
\label{[eq:ns_mass}
\nabla\cdot(\mathrm{u})=0
\end{equation}
\begin{equation}
\label{eq:mom_mass}
\rho\left( \frac{\partial \mathrm{u}}{\partial t} + (\mathrm{u} \cdot \nabla)\mathrm{u} \right) = -\nabla p + \mu \nabla^2 \mathrm{u} + \rho \mathrm{g} + {f}_\gamma
\end{equation}
where u denotes the velocity vector, $\rho$ the mass density, $p$ the static pressure and $g$, the gravitational acceleration. The term $f_{\gamma}$ in momentum equation \ref{eq:mom_mass} evaluates the surface tension forces. The surface tension can be evaluated using different methods in FS3D, which includes continuous surface stress (CSS) model by \textcite{Lafaurie1994}, which has been modified by Potyka et al. \cite{Potyka2023} for evaluating surface tension forces on three-phase flows, hence making it suitable for the current study. The Volume of Fluid (VOF) method by \textcite{hirt1981volume} is applied to distinguish between the different phases by introducing the scalar,~$f$, which represents the volume fraction of the liquid,

\vspace{-1mm}
\begin{equation}
	\label{eq:fdefinition}
	f(\textbf{x},t) = \left\{
	\begin{array}{ll}
		0 & \text{in the disperse phase},\\
		(0,1) & \text{for interfacial cells},\\
		1 & \text{in the continuous phase .}
	\end{array}\right.
\end{equation}
A transport equation for $f$ needs to be solved to obtain the phase distribution over time, 
as per,
\vspace{-1mm}
\begin{equation}
	\label{eq:ftransport}
	\frac{\partial f}{\partial t} + \nabla \cdot \left( f \textbf{u} \right) = 0.
\end{equation}
To increase the accuracy of the corresponding $f$-fluxes, the Piecewise Linear Interface Calculation (PLIC) method by Rider and Kothe \cite{rider1998reconstructing}, which calculates the orientation and position of the interface in each cell is used. A sequential positioning approach by \textcite{Kromer_2022} has been used to reconstruct the interface for 3-phase cells. The density and viscosity in a cell are evaluated in a volume-weighted approach as per the following formulation,
\begin{equation}
\label{eq:property_calculation}
\psi(\textbf{x},t) = \sum_{i=1}^n(\psi_{i}f(\textbf{x},t))
\end{equation}
where $\psi$ represents the physical quantity, $\psi_i$ denotes the physical property value of phase $i$ and $f(x,t)$ denotes the scalar volume of phase $i$ in the cell.

\vspace{-5mm}
\section{Simulation Results}
\label{sec:simulation_results}
\vspace{-4mm}
The first section contains the computational setup. A grid independence study has been conducted to determine the resolution needed for the validation. The validation study has been conducted in two parts, firstly, the effects of viscosity ratio has been studied to assess the changes within impact morphologies. Subsequently, a subcase has been quantitatively validated by comparing the crown heights with experimental results. For the study on the variation of interfacial tension, the identical setup has been used to observe the effects on crown height and spreading diameter. All simulations have been conducted on the HPE Apollo (Hawk) supercomputer.
\vspace{-5mm}
\subsection{Computational Setup}
\label{subsec:computational_setup}
\vspace{-4mm}
\begin{figure}[H]
        \centering
        \includegraphics[width=0.60\linewidth]{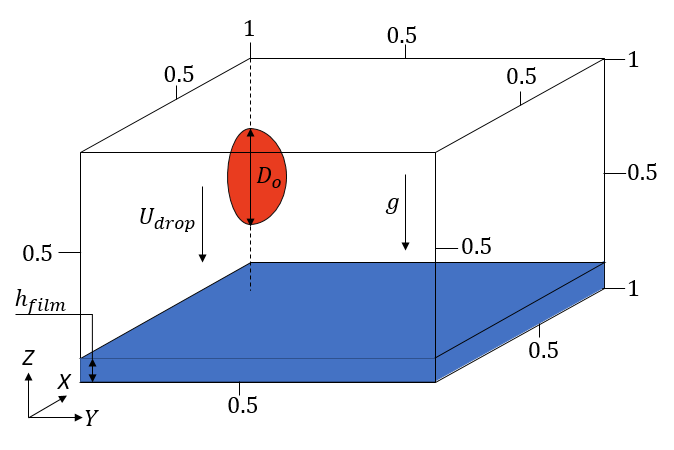}
        \caption{Sketch of the computational domain showing the initial condition and the coordinate system}
        \label{fig:setup_drop_film}
\end{figure}
\begin{table}[ht]
    \caption{Physical properties of air, water and silicon oil \cite{Che2018}}
	\label{tab:grid}
	\centering
	\begin{tabular}{p{2.7cm}p{2.5cm}p{2.5cm}p{1.9cm}}
		\hline\noalign{\smallskip}
		   Material & Density $\rho\;(\mathrm{kg/m^3})$
  & Dynamic viscosity$\mu(\mathrm{Pa\ s})$ & Surface tension $\sigma(\mathrm{mN/m})$ \\
		\noalign{\smallskip}\svhline\noalign{\smallskip}
		  Air & 1.225 & 1.82 x $10^{-5}$ & -- \\
		 Water & 1000 & 0.89 x $10^{-3}$ & 72  \\
            Water + 40\% wt. Glycerol & 1100 & 3.2 x $10^{-3}$ & 69 \\
            Silicon Oil S10 & 930 & 9.3 x $10^{-3}$  & 20 \\
            Silicon Oil S50 & 960 & 4.8 x $10^{-2}$  & 20 \\
		\noalign{\smallskip}\hline\noalign{\smallskip}
	\end{tabular}
	\vspace{-4mm}
\end{table}
The above-mentioned software package, FS3D, has been employed to simulate the drop-film interaction. Figure \ref{fig:setup_drop_film} illustrates the simulation setup, where a droplet of diameter $D_0 = 3.2\,\text{mm}$ impacts a liquid film of thickness ($h$) $0.7\,\text{mm}$ at a velocity of $2.08\,\text{m/s}$. To achieve higher spatial resolution, a quarter-domain representation of the droplet is used, with symmetry planes applied at $Y = 0$ and $X = 1$. The computational domain applies a no-slip boundary condition at the base ($Z = 0$), while continuous boundary conditions are imposed at $X = 0$ and $Y = 1$. In this specific case, a pure water droplet impacts a liquid film of S10 silicone oil, with an interfacial tension of $36\,\text{mN/m}$. The physical properties of the drop and film liquids are summarized in Table \ref{tab:grid}. Non-dimensional numbers used for  further characterization of the impact conditions are the Weber number of the droplet, $We_{\text{drop}}$, calculated as $(\rho U^2 D_0/\sigma)$, is 174; the non-dimensional film thickness, $\delta = h / D_0$, is 0.218; and the Ohnesorge number of the droplet, defined as $\mu_d / \sqrt{\rho_d D_0 \sigma_d}$, is 0.0549. To clearly distinguish between the two fluid phases, a consistent color scheme is used in visualizations—red denotes the droplet liquid, and blue represents the film liquid. The simulation domain spans approximately $3D_0 \times 3D_0 \times 3D_0$.



\vspace{-5mm}
\subsection{Grid Independence Study}
\label{subsec:grid_independence}
\begin{figure}
  \centering
  \begin{minipage}[b]{0.5\textwidth}
    \centering
    \includegraphics[width=\linewidth]{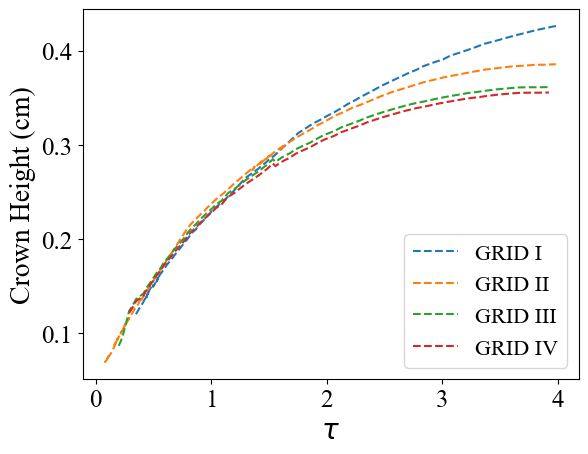}
    \caption{Crown Height development for different grid sizes }
    \label{fig:grid_ind}
  \end{minipage}
  \hfill
  \begin{minipage}[b]{0.45\textwidth}
    \centering
    \begin{tabular}{|c|c|c|}
      \hline
      Name & Total Resolution & Droplet resolution \\
       & & (cells/diameter) \\
      \hline
      GRID I & $512^3$ & 164 \\
      GRID II & $768^3$ & 246 \\
      GRID III & $1024^3$ & 328 \\
      GRID IV & $2048^3$ & 656 \\
      \hline
    \end{tabular}
    \captionof{table}{Specifications of the grid sizes used for grid-dependence study}
    \label{tab:grids_grid_dependency}
  \end{minipage}
\end{figure}

\vspace{-4mm}
To determine the optimal grid resolution for the simulation, four grid sizes were tested, as summarized in Table~\ref{tab:grids_grid_dependency}, with uniform refinement in the X, Y, and Z directions ranging from $512^3$ to $2048^3$ cells. 
\begin{figure}[H]
    \centering
    \begin{subfigure}[b]{0.49\textwidth}
        \includegraphics[width=1\linewidth]{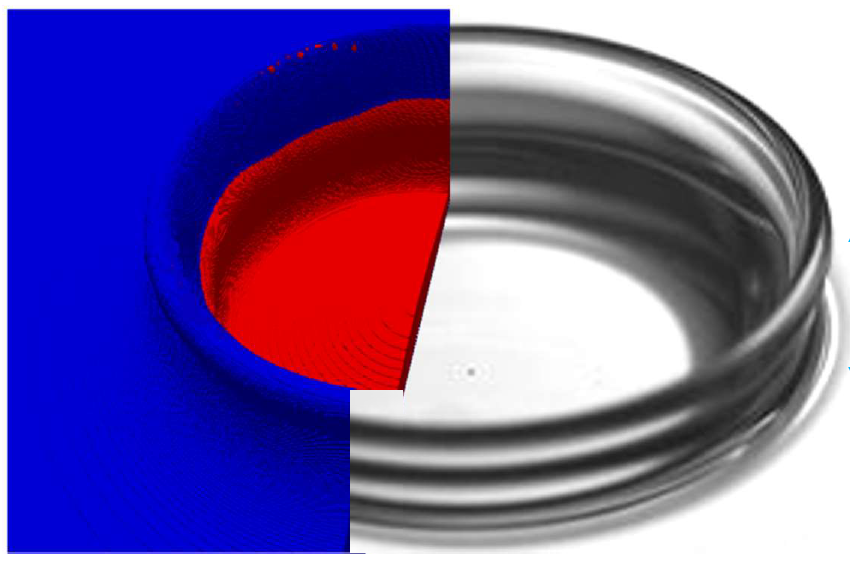}
    \caption{Comparison of water drop over oil film}
    \label{fig:water_oil_1}
    \end{subfigure}
    \hfill
    \begin{subfigure}[b]{0.49\textwidth}
        \includegraphics[width=1\linewidth]{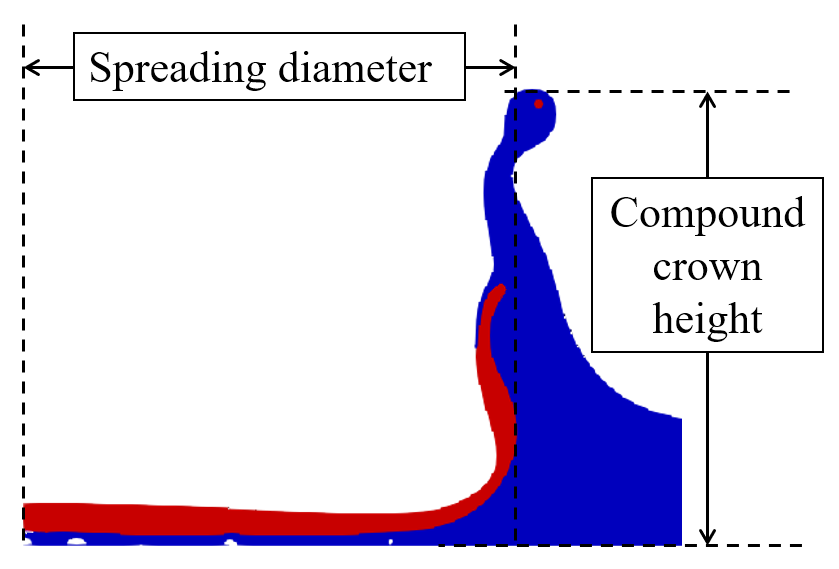}
    \caption{Schematic diagram of a water drop over an oil film}
    \label{fig:water_oil_2}
    \end{subfigure}
    \caption{Water + 40\% glycerol drop with droplet diameter $D_0$=3.06 mm impacting a Silicon Oil S50 film of height 0.7 mm with a velocity of 2.7 m/s.}
    \label{fig:water_oil_phy}
\end{figure}
The non-dimensional form of time has been evaluated as $\tau=\mathrm{tU/D_0}$, where at t = 0 the drop liquid impacts onto the film liquid. All simulations were conducted up to a non-dimensional time of $\tau = 4$ by which the maximum crown height is typically reached. The computational domain was partitioned into $64^3$ cells per MPI process, with the number of compute nodes scaled from 4 for the $512^3$ grid to 64 for the $2048^3$ grid. The simulation with Grid IV required 144 hours on Hawk to complete whereas Grid III required 72 hours on Hawk to run till $\tau = 4$. Results show that crown height flattened by $\tau = 4$, and further refinement beyond $1024^3$ (Grid III) did not affect the outcome as shown in fig. \ref{fig:grid_ind}. Therefore, the Grid III was selected for subsequent analyses, as it provides an optimal balance between accuracy and computational cost. 
\vspace{-5mm}
\subsection{Validation study}
\label{sec:physics_comparison}
\vspace{-4mm}
This section is divided into two parts. The first part presents a qualitative validation study examining the effects of viscosity ratio in FS3D, while the second part offers a quantitative validation focused on a water drop–silicone oil film interaction sub-case.

In the qualitative part, we numerically replicate the experimental setups described by Che and Matar~\cite{Che2018} to investigate the influence of the viscosity ratio during the initial stages of crown formation and propagation. In their experiments, a droplet of S50 silicone oil impacted a water–glycerol (40\%) film at a velocity of $\mathrm{U}_{\mathrm{drop}} = 2.4~\mathrm{m/s}$ and film height $h = 0.7~\mathrm{mm}$. Subsequently, the roles of the droplet and film liquids were reversed under the same impact velocity and film thickness.
\begin{figure}[H]
    \centering
    \begin{subfigure}[b]{0.49\textwidth}
        \includegraphics[width=\linewidth]{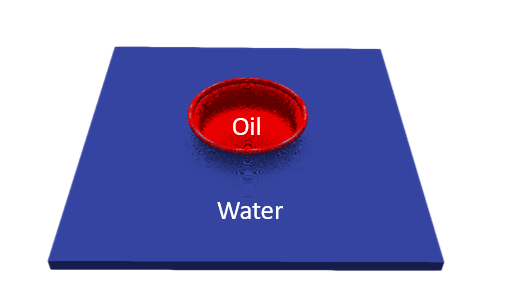}
        \caption{Top view of an silicon oil S50 drop impacting a water+40\%glycerol film}
        \label{fig:Oil_water_2}
    \end{subfigure}
    \hfill
    \begin{subfigure}[b]{0.49\textwidth}
        \includegraphics[width=\linewidth]{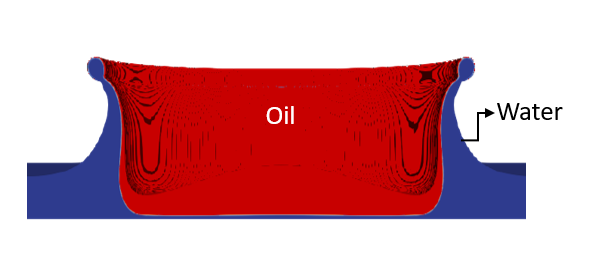}
        \caption{Crown propagation of silicon oil S50 drop impacting a water film which contains 40\%glycerol}
        \label{fig:Oil_water_1}
    \end{subfigure}
    \caption{Silicon Oil S50 drop of droplet diameter $D_o=$ 2.23 mm impacting a water+40\%glycerol solution film of $\delta$=0.219 (h=0.7 mm) with a $u_{\textnormal{drop}}$ of 2.7 m/s.}
    \label{fig:side_by_side}
\end{figure}
In the first configuration—where the silicone oil droplet impacts the water–glycerol film—the drop liquid is observed to spread over the film during crown formation. Based on the surface tensions listed in Table \ref{tab:grid} and an interfacial tension of 24 mN/m (as reported in \cite{Che2018}), the spreading coefficient, defined as ($\sigma_{\mathrm{water-air}} -(\sigma_{\mathrm{oil-air}}+\sigma_{\mathrm{water-oil}})$ , is positive. This positive spreading coefficient suggests that the oil phase preferentially spreads over the water film, consistent with the experimental observations reported by \textcite{Che2018}. The numerical simulations can reproduce similar features, notably the engulfment of the water film by the spreading oil layer during crown formation (fig. \ref{fig:Oil_water_2}). 

When the liquids of drop and film are interchanged, similar to what Che and Matar \cite{Che2018} observed, a compound crown formation is visible where the water drop liquid forms an internal crown owing to the negative spreading factor and adheres to the larger crown formed by the S50 Silicon oil. Numerical investigations also produce results where the separate crowns are visible as shown in fig \ref{fig:water_oil_1}. The compound crown height has been considered to be the crown height, since the internal crown height has not been explicitly evaluated in any experimental findings. The spreading diameter has been considered as the maximum diameter of the drop spreading across the film liquid as shown in fig. \ref{fig:water_oil_2}. In this case, the tension between the drop and film liquid affect the dissipation influencing the spreading, making it favorable for further investigations to study the effects of interfacial tension. 
\begin{figure}[H]
    \centering
    \includegraphics[width=0.5\linewidth]{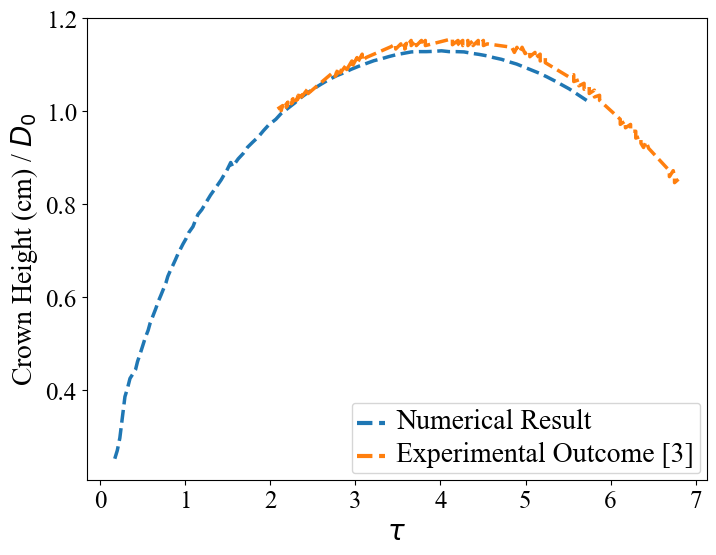}
    \caption{Comparison of experimental crown height for Water drop over Silicon S10 Oil film for a We=174 with numerical simulation results}
    \label{fig:enter-label}
\end{figure}
For further investigation the grid III, which was found suitable as per the grid independence study, has been run further till $\tau= 6$ with the material properties and impact conditions stated in section \ref{subsec:computational_setup}. The evolution of the crown height has been compared with the experimental crown heights and the results provide a good quantitative match during the crown evolution with a variation score ($\mathrm{R^2}$) of 0.98 till $\tau= 6$. The R\textsuperscript{2} score was calculated as 
$\mathrm{R^2} = 1 - \frac{\sum (y_{\text{exp}} - y_{\text{num}})^2}{\sum (y_{\text{exp}} - \bar{y}_{\text{exp}})^2}$ comparing the variance of residuals between experimental crown height values \( y_{\text{exp}} \) and numerical predictions \( y_{\text{num}} \) to the total variance of \( y_{\text{exp}} \), where $\bar{y}_{\text{exp}}$ is the mean value of \( y_{\text{exp}} \).

A close look at the morphology, as shown in fig. \ref{fig:water_oil_2}, reveals that there is no rim breakup occurring at the highest position of crown height and no secondary droplets are visible till the time step it has been evaluated.  
\vspace{-5mm}
\subsection{Role of interfacial tension}
\label{subsec:comparison_interfacial_tension}
\begin{figure}
    \centering
    \includegraphics[width=1\linewidth]{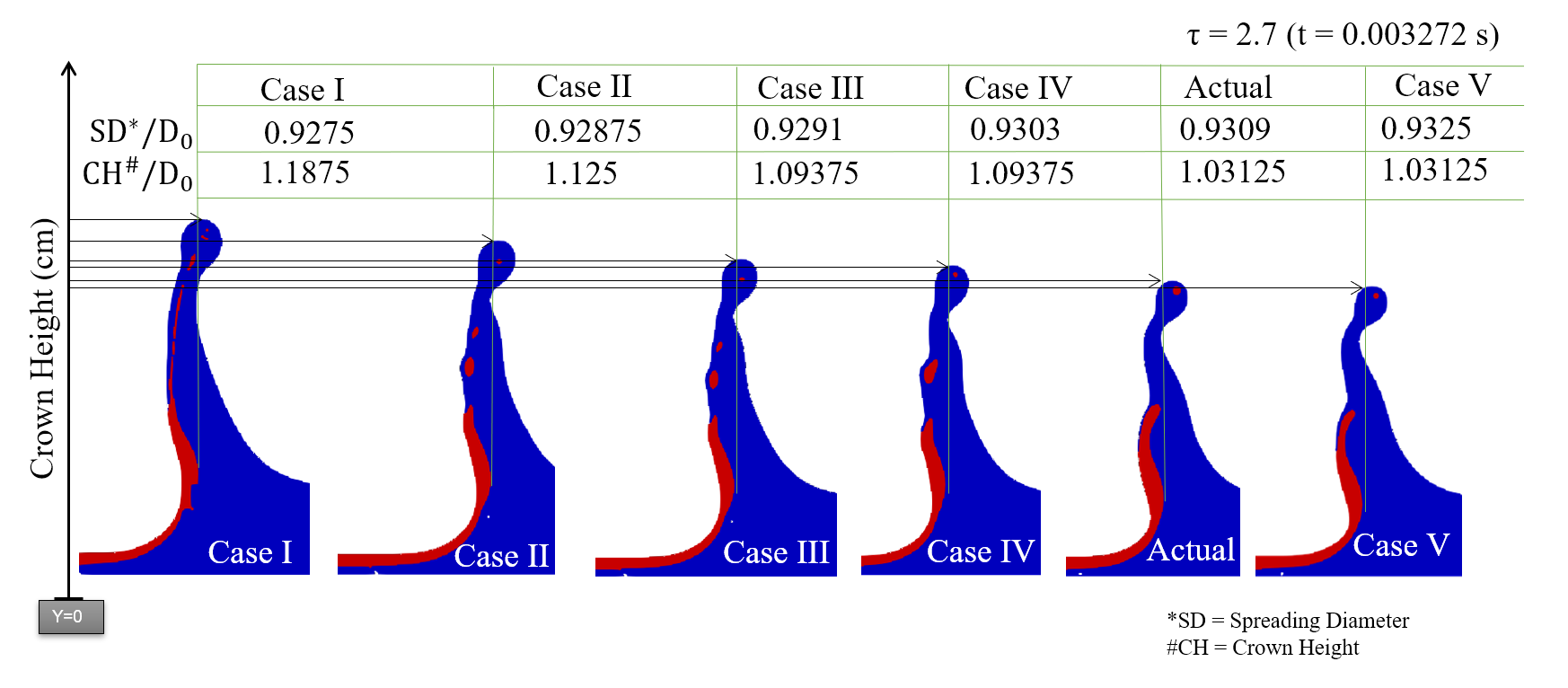}
    \caption{Variation in the crown morphology with increasing interfacial tension}
    \label{fig:chg_int_ten}
\end{figure}

\vspace{-4mm}
After proving a good morphological agreement, the influence of interfacial tension is studied numerically. In order to investigate the role of interfacial tension, grid III is used. The impact conditions and the material properties are kept identical as the validation case. Five cases are considered where the interfacial tension is changed to values in multiples of 10 from 0 till 40. The comparisons have been done at $\tau=2.7$, during the initial phase of crown propagation. \\
The results as per fig. \ref{fig:chg_int_ten} reveal that as the interfacial tension increases the crown height decreases, considering that the surface energy increases with increasing interfacial tension. There is higher dispersion of the drop liquid within the film liquid in Case I where the interfacial tension is zero, representing a scenario of "miscible-like" liquid interaction as we see greater volume of drop liquid reaching the compound crown rim. This can be accounted to the lower amount of kinetic energy conversion to surface energy as the surface energy is a function of the partial surface tensions evaluated algebraically from the interfacial tension \cite{Potyka2024}. The variation in spreading diameter of the drop liquid is comparatively negligible as plotted in the Table \ref{tab:error}. 
\begin{table}
    \caption{Deviation \% in crown height and spreading diameter with different interfacial tension}
    \label{tab:error}
    \centering
    \begin{tabular}{
        >{\centering\arraybackslash}p{1.5cm}
        >{\centering\arraybackslash}p{3.2cm}
        >{\centering\arraybackslash}p{3.8cm}
        >{\centering\arraybackslash}p{3.8cm}}
        \hline\noalign{\smallskip}
        Case & Interfacial tension (mN/m) & \% Deviation for Crown Height & \% Deviation for Spreading Diameter \\
        \noalign{\smallskip}\hline\noalign{\smallskip}
        Case I   & 0  & 15 & 0.37 \\
        Case II  & 10 & 9 & 0.24 \\
        Case III & 20 & 6 & 0.20 \\
        Case IV  & 30 & 6 & 0.06 \\
        Case V   & 40 & -- & 0.16 \\
        \noalign{\smallskip}\hline
    \end{tabular}
    \vspace{-4mm}
\end{table}
Although the crown height and spreading diameter deviations are not significant as compared to the variation in the interfacial tension, the internal dynamics is quite different as visible in fig. \ref{fig:chg_int_ten}. There is pinching and secondary droplet formation at the rim of the internal crown by the drop liquid for cases I, II, III and IV. This effect is unique and needs further investigation, in order to check if further increase in resolution affects the results, or if such effects can be experimentally observed for liquids with lower interfacial tensions. As a part of the future task, varying the Weber number or incorporating the interfacial tension within the Weber number can be interesting to check how varying velocities of drop liquid can lead to different outcomes. 

\vspace{-5mm}
\section{Computational Performance}
\label{sec:performance}
\vspace{-4mm}
Supercomputer vendors constantly strive to improve the performance and efficiency of their systems.
A new direction is to use accelerators like GPUs to offload compute- and memory-intensive tasks, even for scientific computing.
The new system Hunter at the High-Performance Computing Center Stuttgart (HLRS) follows this trend
and uses 4 AMD Instinct™ MI300A accelerated processing units (APUs) per node,
where each APU consists of 24 CPU cores and 228 compute units (accelerated part).
While FS3D was optimized for vector computers back in the days,
and later ported to efficiently use highly parallel CPU clusters,
the new system at HLRS now requires porting the code to efficiently utilize the new APU architecture.

In the following, we will describe the general porting strategy,
the idea of using a memory pool and performance tests for the new system.
With the performance tests, the efficient use of the APU is investigated and
weak and strong scaling is shown.
\vspace{-5mm}
\subsection{Offloading strategy}
\vspace{-4mm}
As an offloading strategy, we use the OpenMP pragma-based approach.
This has the advantage that, in most cases, only OpenMP directives need to be added to the code,
and, thanks to the unified shared memory (USM) of the APU, no explicit data transfers are needed.
This results in moderate code changes necessary for porting
and allows the code to remain portable.
An example directive could be:\\
\code{!\$OMP TARGET TEAMS DISTRIBUTE PARALLEL DO COLLAPSE(3)\\
DEFAULT(NONE) SHARED(a,in) \-PRIVATE(i,j,k)}\\
before 3 nested loops.
Here the \code{TARGET TEAMS DISTRIBUTE PARALLEL DO} enables the offloading of the code,
and the \code{COLLAPSE(3)} directive allows the compiler to parallelize over all 3 loops.
The \code{DEFAULT(NONE)} directive ensures that all variables are explicitly declared,
avoiding different default values for different compilers and reducing programming errors.
A further advantage is that the exact same code can be used for runs on CPU-only systems
and on accelerated systems when using USM.
While for parallelization on APU level the above described OpenMP directives are used,
for multiple APUs, FS3D uses domain decomposition with Message Passing Interface (MPI) for parallelization.
This results in a hybrid parallelization scheme,
where one MPI rank is assigned to one APU,
and OpenMP parallelization is used to distribute the work within one APU.

While the major part of the used code is ported,
note that it is not yet fully ported and optimized for the current system.
There are some parts of the code taking significant time, but have not been offloaded yet.
Further performance improvements, especially at APU level, are expected.
\vspace{-5mm}
\subsection{Memory pool}
\vspace{-4mm}
The unified shared memory (USM) of the APU has the advantage that no data transfers between CPU and GPU are needed,
which not only simplifies the code but also improves performance.
Tests showed that frequent allocations on the CPU using \code{ALLOCATE()} and subsequent computation on the accelerated part
show some performance penalty compared to computation on already allocated and reused memory%
\footnote{See AMD training material:
\url{https://github.com/amd/HPCTrainingExamples/tree/main/Rocprofv3/OpenMP/Allocations_and_MemoryPool_MI300A/Fortran}}.
To avoid this penalty, Umpire \cite{umpire}, a memory management API for non-uniform memory access (NUMA) and GPU architectures,
can be used to manage the memory.
Umpire allows the creation of a memory pool so that memory is allocated once and can be reused and reassigned without
the need for repeated allocations.
Performance tests are shown in the following subsection.
\vspace{-5mm}
\subsection{Scaling and performance}
\vspace{-4mm}
For scaling and performance tests, direct numerical simulations (DNS) of an oscillating droplet is carried out.
There the full 3D Navier-Stokes equations are solved and other physical effects like surface tension,
which are fundamental for multiphase flows, are included.
It is the same physical setup as in \cite{stober2023dns}
to enable a direct comparison between the previous HLRS supercomputer, Hawk,
and their current system, Hunter. 

For some parts of the code, where OpenMP parallelization is used
but are not offloaded to the GPU,
24 OpenMP threads are used.
\vspace{-1mm}
\subsubsection{Scaling and performance for constant compute power}
\begin{figure}
    \centering
    \smallskip

    \begin{subfigure}[t]{0.49\textwidth}
        \centering
    \begin{tikzpicture}
        \begin{axis}[
            width=0.98\textwidth,
            height=0.28\textheight,
            name=ax1,
            xlabel={cells per APU},
            ylabel style={align=center},
            ylabel={average time per cycle [\SI{}{\s}]},
            ymode=log,
            xmode=log,
            log ticks with fixed point,
            legend style={
                font=\footnotesize,
                legend columns=2,
                draw=gray,
                at={(1.05,1.02)},
                anchor=south
            },
            mark options=solid,
            xtick={64^3, 128^3, 256^3, 512^3}, 
            xticklabels={$64^3$, $128^3$, $256^3$, $512^3$} 
        ]

            \addplot[color=color1, mark=*]%
            table[x=totalCellsPerAPU, y=avgTimePerCycle, col sep=comma] {csv_files/umpire1APU.csv};
            \addlegendentry{Umpire, 1 APU}

            \addplot[color=color2, mark=square*]%
            table[x=totalCellsPerAPU, y=avgTimePerCycle, col sep=comma] {csv_files/umpire2Nodes.csv};
            \addlegendentry{Umpire, 2 Nodes}

            \addplot[color=color1, mark=triangle*, dashed]%
            table[x=totalCellsPerAPU, y=avgTimePerCycle, col sep=comma] {csv_files/noUmpire1APU.csv};
            \addlegendentry{No Umpire, 1 APU}

            \addplot[color=color2, mark=diamond*, dashed]%
            table[x=totalCellsPerAPU, y=avgTimePerCycle, col sep=comma] {csv_files/noUmpire2Nodes.csv};
            \addlegendentry{No Umpire, 2 Nodes}

            \addplot[color=color1, mark=x, dotted]%
            table[x=totalCellsPerAPU, y=avgTimePerCycle, col sep=comma] {csv_files/CPU1APU.csv};
            \addlegendentry{CPU only, 1 APU}

            \addplot[domain=64^3:512^3, samples=500, color=gray, densely dotted] {7/512^3*x};
            \addlegendentry{perfect scaling}
        \end{axis}
    \end{tikzpicture}
    \caption{Average time per cycle.}
    \label{fig:time_per_cycle}
    \end{subfigure}
    \hfill
    \begin{subfigure}[t]{0.49\textwidth}
        \centering
    \begin{tikzpicture}

        \begin{axis}[
            at={(ax1.south east)},
            xshift=2cm,
            width=0.98\textwidth,
            height=0.28\textheight,
            xlabel={cells per APU},
            ylabel style={align=center},
            ylabel={efficiency [-]},
            xmode=log,
            log ticks with fixed point,
            legend to name=legend2,
            grid=none,
            mark options=solid,
            xtick={64^3, 128^3, 256^3, 512^3}, 
            xticklabels={$64^3$, $128^3$, $256^3$, $512^3$} 
        ]

            \addplot[color=color1, mark=*]%
            table[x=totalCellsPerAPU, y=efficiency_toMax, col sep=comma] {csv_files/umpire1APU.csv};
            \addlegendentry{Umpire 1 APU}

            \addplot[color=color2, mark=square*]%
            table[x=totalCellsPerAPU, y=efficiency_toMax, col sep=comma] {csv_files/umpire2Nodes.csv};
            \addlegendentry{Umpire 2 Nodes}

            \addplot[color=color1, mark=triangle*, dashed]%
            table[x=totalCellsPerAPU, y=efficiency_toMax, col sep=comma] {csv_files/noUmpire1APU.csv};
            \addlegendentry{No Umpire 1 APU}

            \addplot[color=color2, mark=diamond*, dashed]%
            table[x=totalCellsPerAPU, y=efficiency_toMax, col sep=comma] {csv_files/noUmpire2Nodes.csv};
            \addlegendentry{No Umpire 2 Nodes}

            \addplot[color=color1, mark=x, dotted]%
            table[x=totalCellsPerAPU, y=efficiency_toMax, col sep=comma] {csv_files/CPU1APU.csv};
            \addlegendentry{CPU 1 APU}

            \addplot[domain=64^3:512^3, samples=500, color=gray, densely dotted] {1};
            \addlegendentry{perfect scaling}
        \end{axis}
    \end{tikzpicture}
    \caption{Efficiency relative to the fastest $512^3$ cells per APU setup (Umpire, 1 APU).
    }
    \label{fig:efficiency}
    \end{subfigure}
    \caption{Scaling and performance for constant compute power.
    Performance comparisons between Umpire memory pool and standard Fortran allocation
    are shown for both single-APU and 2-node (8 APUs) setup.
    The 2-node setup has always 8 times more total cells than the 1 APU setup.
    As comparison, the CPU-only setup is shown, which uses the CPU part of 1 APU with 16 MPI ranks.
    }
    \label{fig:const_compute_power}
\end{figure}
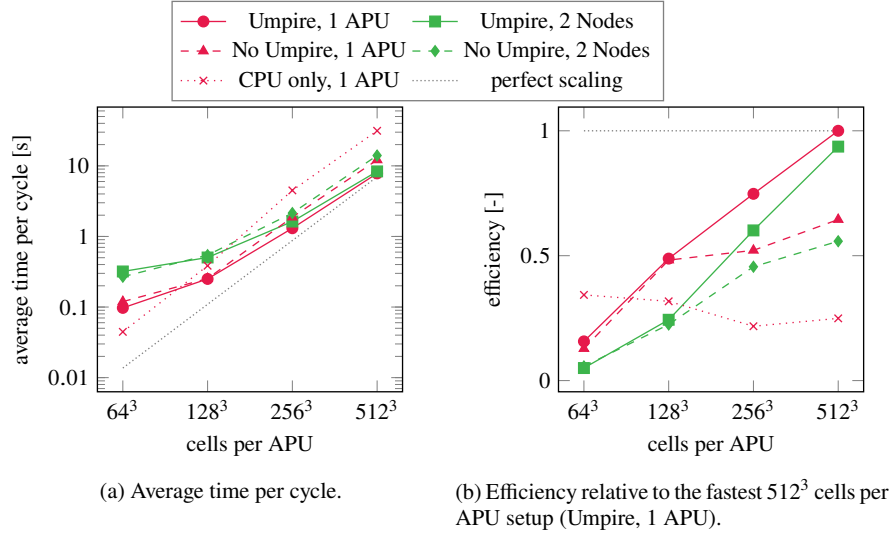

\vspace{-3mm}
Figure \ref{fig:const_compute_power} shows the performance with constant compute power,
but decreasing cell count (decreasing workload) per APU. This can help answer the question of how large the problem size per APU should
be to work efficiently on the APU, and provides insights into balancing resource efficiency and time-to-solution. For FS3D a typical kernel operates on the number of cells per APU.
Four different setups are used that run on the GPU part of the APU.
Two setups use the Umpire memory pool and the other two setups use standard Fortran memory allocation.
This allows a comparison of the performance of allocations using Umpire
and allocations without Umpire.
Additionally, in two setups, 1 APU and 1 MPI rank is used,
while in the other two setups, 2 nodes with 4 APUs each (8 MPI ranks) are used.
As the cells per APU are the same for the 1 APU and 2 nodes setups,
this results in $512^3$ cells for 1 APU and $1024^3$ cells for 2 nodes
for the maximum cell count per APU of $512^3$.
$512^3$ cells for 1 APU is the maximum that fits in the memory for this physical setup.
The fifth setup uses only the CPU part of 1 APU with 16 MPI ranks,
no OMP, but cache line optimizations as in \cite{potyka2022HLRS}.
This setup is used to quantify the performance improvements due to offloading.

The average time per cycle is shown in Figure \ref{fig:time_per_cycle}.
The theoretical perfect scaling is shown as a gray dotted line.
The double logarithmic plot shows the expected decreasing time per cycle
for decreasing cells per APU across all cases.
At $512^3$ cells per APU,
the best performance is achieved with the Umpire memory pool and 1 APU,
where the time per cycle is \SI{7.8}{\s} (or 460 cycles per hour (CPH)).
For $512^3$ cells per APU, compared to without Umpire,
the time per cycle decreases with Umpire
by \SI{4.3}{\s} (or by \SI{36}{\percent}) for the 1 APU setup
and by \SI{5.6}{\s} (or by \SI{40}{\percent}) for the 2 nodes setup.
This demonstrates that using the Umpire memory pool improves performance
for larger cell counts per APU.
For lower cell counts per APU, the performance is similar when using Umpire and
without Umpire.
In the case of 2 nodes,
the performance is always worse than for 1 APU (when using also the GPU part) at the same cell count per APU.
This is expected, as there is additional time spent for communication.
At $512^3$ cells per APU, the time per cycle increases by
\SI{0.5}{\s} (or by \SI{7}{\percent}) for the 2 nodes setup with Umpire.
The time per cycle for the CPU-only setup is \SI{31}{\s} (or 115 CPH) for $512^3$ cells per APU,
which is about 4 times slower than the 1 APU setup with Umpire.

The efficiency is shown in Figure \ref{fig:efficiency},
and is calculated as the theoretical time per cycle derived from the time per cycle
of the fastest setup (Umpire, 1 APU) at $512^3$ cells per APU,
divided by the time per cycle of the current cell count per APU.
The efficiency is always worse for the 2 nodes setup compared to the 1 APU setup,
and always better with Umpire compared without Umpire.
Overall, the efficiency decreases with decreasing cell count per APU when using the GPU part of the APU,
but increases slightly for the CPU-only setup.
The presumed reason for the decreasing efficiency of the 1 APU setup when using the GPU part of the APU
is the decreased kernel size and the presumed reason for the increasing efficiency of the CPU-only setup
is the better cache usage.

The 4 times higher performance of the 1 APU setup with Umpire compared to the CPU-only setup
(at $512^3$ cells per APU) demonstrates the success of the porting efforts.
It is also shown that larger cell counts per APU are beneficial for efficiency,
however, at APU level, smaller cell counts per APU improves time-to-solution,
without losing too much efficiency.
The next section shows the scaling for multiple APUs.
There the strong scaling investigates if reduced cell counts per APU
actually improve the time-to-solution.

\subsubsection{Strong and weak scaling}
\vspace{-3mm}
Figure \ref{fig:StrongScalingSeconds} shows the strong scaling of FS3D for a $512^3$ cells simulation,
starting with 1 APU and 1 MPI rank and progressively increasing to 256 APUs and 256 MPI ranks.
For 256 APUs, the number of cells per APU is only $64^2\cdot128$.
The strong scaling shows good scaling for up to 4 APUs, but negligible speedup starting at 16 APUs.
Comparing this to Figure \ref{fig:time_per_cycle}, where there is some speedup even for $64^3$ cells per APU,
reveals that the not so good strong scaling is not solely due to the smaller kernel sizes.
Taking the example of 64 APUs, the cells per APU are $128^3$ and the time per cycle is \SI{1}{\s}.
Looking into Figure \ref{fig:time_per_cycle}, the time per cycle for also $128^3$ cells on 1 APU is \SI{0.25}{\s},
which indicates the major part of the efficiency loss is not due to smaller kernel sizes,
but rather due to communication latency, load imbalances, different switch-levels for the Multigrid
or other effects.

Figure \ref{fig:WeakScalingSeconds} shows the weak scaling of FS3D for
$512^3$ and for $256^3$ cells per APU.
The physical setup is the same as for the previous section,
but the number of cells per APU is kept constant and the number of APUs is increased.
For $512^3$ cells per APU and the maximum number of APUs (512),
the total number of cells is $4096^3$,
a problem size that was not possible before.
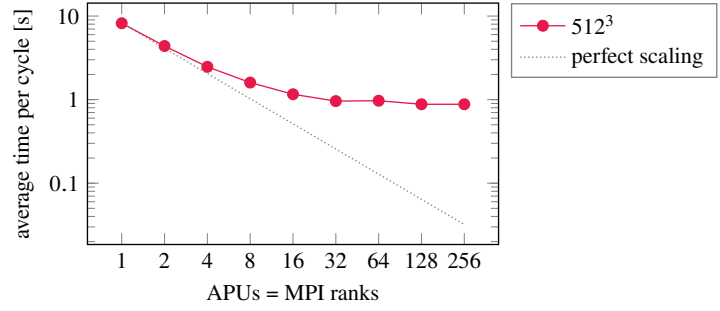
\begin{figure}
    \centering
    \begin{tikzpicture}
        \begin{axis}[
            width=0.6\textwidth,
            height=0.25\textheight,
            xlabel={APUs = MPI ranks},
            ylabel style={align=center},
            ylabel={average time per cycle [\SI{}{\s}]},
            ymode=log,
            xmode=log,
            log ticks with fixed point,
            legend pos=outer north east,
            legend style={font=\footnotesize, draw=gray},
            legend cell align={left},
            grid=none,
            mark options=solid,
            xtick={1, 2, 4, 8, 16, 32, 64, 128, 256}, 
            xticklabels={1, 2, 4, 8, $16$, 32, 64, 128, 256} 
        ]

            \addplot[color=color1, mark=*,/pgf/number format/read comma as period]%
            table[x=ranks, y=time/cycle, col sep=semicolon] {csv_files/Strong_Scaling_512x512x512.csv};
            \addlegendentry{$512^3$}

            \addplot[domain=1:256, samples=500, color=gray, densely dotted] {8.21/x};
            \addlegendentry{perfect scaling}
        \end{axis}
    \end{tikzpicture}
    \caption{Strong scaling of FS3D for a fixed problem size of $512^3$ cells.
    }
    \label{fig:StrongScalingSeconds}
\end{figure}
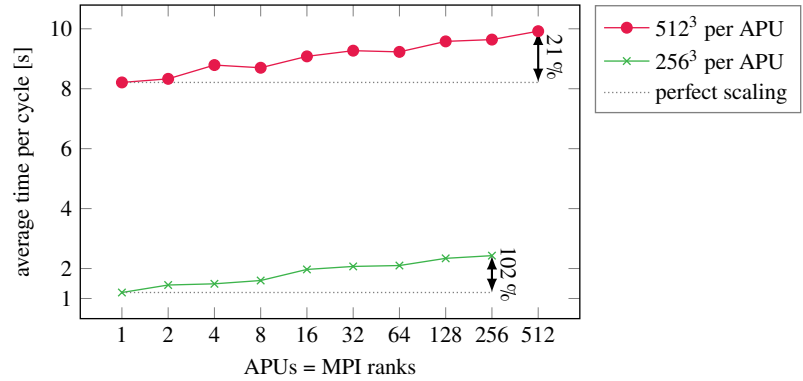
\begin{figure}
    \centering
    \begin{tikzpicture}
        \begin{axis}[
            width=0.7\textwidth,
            height=0.3\textheight,
            xlabel={APUs = MPI ranks},
            ylabel style={align=center},
            ylabel={average time per cycle [\SI{}{\s}]},
            xmode=log,
            log ticks with fixed point,
            ytick={1, 2, 4, 6, 8, 10},
            legend pos=outer north east,
            legend style={font=\footnotesize, draw=gray},
            legend cell align={left},
            grid=none,
            mark options=solid,
            xtick={1, 2, 4, 8, 16, 32, 64, 128, 256, 512}, 
            xticklabels={1, 2, 4, 8, $16$, 32, 64, 128, 256, 512} 
        ]

            \addplot[color=color1, mark=*,/pgf/number format/read comma as period]%
            table[x=ranks, y=time/cycle, col sep=semicolon] {csv_files/Weak_Scaling_512x512x512_perAPU.csv};
            \addlegendentry{$512^3$ per APU}

            \addplot[color=color2, mark=x,/pgf/number format/read comma as period]%
            table[x=ranks, y=time/cycle, col sep=semicolon] {csv_files/Weak_Scaling_256x256x256_perAPU.csv};
            \addlegendentry{$256^3$ per APU}

            \addplot[domain=1:512, samples=500, color=gray, densely dotted] {8.21};
            \addplot[domain=1:256, samples=500, color=gray, densely dotted] {1.2};
            \addlegendentry{perfect scaling}

            \draw[latex-latex, thick] (axis cs:512, 8.21) -- (axis cs:512, 9.92) node[midway, above, sloped, rotate=180] {$\SI{21}{\percent}$};
            \draw[latex-latex, thick] (axis cs:256, 1.2) -- (axis cs:256, 2.43) node[midway, above, sloped, rotate=180] {$\SI{102}{\percent}$};
        \end{axis}
    \end{tikzpicture}
    \caption{Weak scaling of FS3D for $512^3$ and $256^3$ cells per APU.
    The numbers on the arrows show the relative increase of the time per cycle.
    }
    \label{fig:WeakScalingSeconds}
\end{figure}

For 512 APUs the time per cycle increases by
\SI{1.7}{\s} (or by \SI{21}{\percent}) compared to 1 APU.
For $256^3$ cells per APU and the maximum number of APUs (256),
the total number of cells is $1024\cdot2048^2$, and the time per cycle increases by
\SI{1.2}{\s} (or by \SI{102}{\percent}) compared to 1 APU.
On Hawk \cite{stober2023dns},the weak scaling showed decreasing performance for increasing cores within one node, then close to constant performance from 1 up to 128 nodes (about \SI{3.6}{\s} per cycle),
but sudden drops in performance for 256 nodes (resulting in $2048^3$ cells total at 256 nodes).
Here time per cycle increases about linearly with the number of APUs for both setups,
without reaching a limit in number of APUs, where the performance drops.
This is a significant improvement compared to Hawk,
where the performance dropped before $2048^3$ cells total.
Here even $4096^3$ cells total are possible,
which is a significant increase in problem size.

Notably, the slope of the time per cycle is similar for both setups,
even though the number of cells per APU is 8 times higher for $512^3$ cells per APU.
This suggests that the increased time per cycle for higher APU counts
is not due to communication bandwidth,
but rather due to communication latency, load imbalances, different switch-levels for the Multigrid
or other effects.
Further investigations are needed to identify the exact reasons for the increased time per cycle.

\vspace{-4mm}
\section{Conclusions}
\vspace{-4mm}
The early phase of crown formation in a drop–film interaction involving immiscible fluids has been numerically investigated using the multiphase DNS solver FS3D. Due to discrepancies in reported interfacial tension values for the same fluid pair, a numerical sensitivity study was conducted to assess the impact of this parameter, using a code previously validated against experimental data. Validation was performed by comparing the compound crown height resulting from a water droplet impacting a silicon S10 oil film, with good agreement observed.

For the sensitivity analysis, all material properties and impact conditions were held constant while varying the interfacial tension from 0 to 40 mN/m. Results indicate a minor change in crown height with increasing interfacial tension, while the spreading diameter remains largely unaffected. Although these variations in overall splash morphology appear small, the internal crown structure exhibits differences that may be influenced by grid resolution. A pinch-off effect at the internal crown rim is observed with secondary droplet formation at lower interfacial tension values. Future studies leveraging higher-resolution simulations, enabled by improved hardware resources, can offer deeper insights into these internal dynamics. 

The new system at HLRS uses an APU based architecture.
For an efficient use of this new architecture major parts of the code FS3D have been successfully offloaded to the GPU using OpenMP.
The performance of the code is improved by implementing the Umpire memory pool,
which enables memory reuse and eliminates frequent allocations.
Scaling and performance testing was first conducted for constant compute power,
with decreasing cell count per APU.
Showing that $512^3$ cells per APU (maximum that fits in memory)
provides optimal resource efficiency.
While for $256^3$ cells at 1 APU still maintains good resource efficiency,
the efficiency declines for lower cell counts per APU.
The weak scaling analysis shows promising results for $512^3$ cells per APU,
suggesting that simulations with $4096^3$ cells could be feasible on 512 APUs.
However, weak scaling tests with $256^3$ cells per APU and strong scaling results
indicate that scaling is limited by communication latency, load imbalances,
or other effects, rather than by communication bandwidth or small kernel sizes.
This needs further investigation.
Further performance improvements will be possible by, e.g. offloading additional parts,
include conditional offloading or using communication-hiding techniques.
The results demonstrate that the porting of FS3D to the GPU using OpenMP
is successful and provides significant performance improvements.

\newenvironment{acknowledgments}%
{\null\begin{center}%
 	\bfseries Acknowledgments\end{center}}%
{\null}
\vspace{-5mm}
\begin{acknowledgments}
The authors kindly acknowledge the {\it High Performance Computing Center Stuttgart} (HLRS) for support and supply of computational resources on the HPE Apollo (Hawk) and HPE Cray EX4000 (Hunter) platform under the Grant No. {FS3D/11142}. Additionally, R.D acknowledges the financial support through the project number 270852890 DROPIT/GRK 2160/2 and D.G acknowledges the support by Project ID 498601949 – TRR 364 funded by the Deutsche Forschungsgemeinschaft (DFG, German Research Foundation). 

The authors gratefully acknowledge the support by AMD (Johanna Potyka), HPE (Andreas Prell, Christian Simmendinger) and HLRS for their support in the porting of FS3D to the APU architecture.\\
\end{acknowledgments}

\vspace{-8mm}

\printbibliography

\eject
\end{document}